# The speed of light under the IST and Lorentz Transformations

Chandru Iyer[1]

[1]Techink Industries, C-42, phase-II, Noida, India – 201305
[1]e-mail: chandru_i@yahoo.com

**Abstract:** We expand the IST transformation to three-dimensional Euclidean space and derive the speed of light under the IST transformation. The switch from the direction cosines observed in K to those observed in K' is surprisingly smooth. The formulation thus derived maintains the property that the round trip speed is constant. We further show that under the proper synchronization convention of K', the one-way speed of light becomes constant.

## 1. Introduction

The IST transformation has been recently discussed in [1, 2]. It is based on constructing the transformation of event coordinates between two inertial frames as a composition of observed physical phenomena. As stated in [1], the *IST transformation* (inertial-synchronized-Tangherlini) was named inertial transformation by Selleri [3], and obtained by Tangherlini [4].

The observations of an inertial frame K, about another inertial frame K' are, as described in [2]:

i) Every object in K' is moving at a speed $v$ along the line of motion,

ii) Objects in K' are contracted by a factor $\sqrt{1-v^2/c^2}$,

iii) A particular clock of K' runs slow by the factor $\sqrt{1-v^2/c^2}$, and

iv) The clocks of K' that are assumed to be synchronous by K' are not synchronous as observed by the clocks of K.

The IST transformation interprets the first three of the four observations as physical effects. The fourth observation – that spatially separated clocks of K' that are believed to be synchronous by observers in K' are essentially observed to be asynchronous by observers in K – is not construed to be a physical effect (by observers in K), as synchronization procedures can be difficult to develop [5, 6] and intuitively, the fact that two spatially separated clocks are asynchronous may be due to a faulty procedure of synchronization and may not imply any deviation in the physical phenomena such as the clock ticks.

It may be pointed out that for observing the first three 'physical phenomena' in K', observers in K trust that their spatially separated clocks are synchronous. Observers in K believe that the procedures adopted by them to synchronize spatially separated and co-moving clocks in their inertial frame are reliable. That the same procedures adopted by observers in K' produced a different set of synchronization is essentially due to the relative motion between K and K'. The disagreement between K and K' on synchronicity of spatially separated clocks is mutual. Under the synchronization convention adopted by K', observers in K' observe identical physical effects about objects of K. Thus the Lorentz transformation which is composed [2] of all the four components in the order as listed above produces a resultant transformation that is symmetric and maintains the equivalence of any two inertial frames.

However, under the proper synchronization convention of K, observers in K do not consider the observed asynchronization of spatially separated clocks of K' as a physical effect. Thus observers in K consider only the first three physical effects listed above to compose the transformation of event coordinates between K and K'. The resultant transformation, composed of only the first three physical effects listed above is known as the IST transformation [1].

In this note we expand the IST transformation to three-dimensional Euclidean space and derive the speed of light under the IST transformation in three-dimensional space. The switch from the direction cosines of K to those of K' is surprisingly smooth and the formulation thus derived maintains the property that the round trip speed is constant. We further show that under the proper synchronization convention of K', the one-way speed of light becomes constant.

## 2. IST transformation

The IST transformation has been discussed [1] in the literature as that which maintains the synchronicity convention of a given frame K and transforms the event coordinates from K to K'. It is defined in one-dimensional space as

$$x' = (x-vt)\gamma$$
$$t' = t/\gamma$$

where the relative motion is along the x, x' axis and $\gamma = \dfrac{1}{\sqrt{1-v^2/c^2}}$.

The term –vt, in the first equation signifies relative motion. For any particular point object located in K' at x' = a, x = (a/$\gamma$) + vt and (dx/dt) = v, signifying a relative motion at a constant speed of v along the x-axis, the line of motion. The factor $\gamma$ in the first equation signifies the contraction of rulers in K' along the line of motion. Due to this contraction, distances are measured by K' to be increased by a factor $\gamma$. The factor $\gamma$ in the second equation signifies the slow running of any particular clock of K'.

In three-dimensional space the two additional equations y' = y; z' = z are in order. Note that there is no observed contraction of rulers of K' in the directions perpendicular to the line of motion. Thus y' and z' remain identical to y and z.

Expressed in the matrix notation the transformation becomes.

$$\begin{pmatrix} x' \\ y' \\ z' \\ t' \end{pmatrix} = \begin{pmatrix} \gamma & 0 & 0 & -v\gamma \\ 0 & 1 & 0 & 0 \\ 0 & 0 & 1 & 0 \\ 0 & 0 & 0 & 1/\gamma \end{pmatrix} \begin{pmatrix} x \\ y \\ z \\ t \end{pmatrix} \quad (1)$$

where ( x, y, z, t) are the event coordinates in frame K and x', y', z', t', are the event coordinates in frame K'.

## 3. Speed of Light under IST transformation

For a light ray observed by K, to be propagating along an arbitrary line with angle $\phi$ with z-axis and with its projection on the x-y plane subtending an angle $\theta$ with the x-axis, we have the following event generated at any time t (as observed by K).

$x = ct \sin\phi \cos\theta$
$y = ct \sin\phi \sin\theta$
$z = ct \cos\phi$

The above event is transformed by the IST transformation to

$$\begin{pmatrix} \gamma & 0 & 0 & -v\gamma \\ 0 & 1 & 0 & 0 \\ 0 & 0 & 1 & 0 \\ 0 & 0 & 0 & 1/\gamma \end{pmatrix} \begin{pmatrix} ct \sin\phi \cos\theta \\ ct \sin\phi \sin\theta \\ ct \cos\phi \\ t \end{pmatrix} = \begin{pmatrix} x' \\ y' \\ z' \\ t' \end{pmatrix}$$

Noting t' =t/$\gamma$ , we get

$$\begin{pmatrix} x' \\ y' \\ z' \\ t' \end{pmatrix} = \begin{pmatrix} c\gamma^2 t' \sin\phi \cos\theta - v\gamma^2 t' \\ c\gamma t' \sin\phi \sin\theta \\ c\gamma t' \cos\phi \\ t' \end{pmatrix} \quad (2)$$

The components of the observed velocity become

$$C_{x'}^+ = c\gamma^2 \sin\phi \cos\theta - v\gamma^2 \quad (3)$$

$$C_{y'}^+ = c\gamma \sin\phi \sin\theta \quad (4)$$

$$C_{z'}^+ = c\gamma \cos\phi \quad (5)$$

Squaring and adding the component velocities and then taking the square root, we get

$$C^+ = c\gamma^2 \sqrt{(\sin\phi\cos\theta - (v/c))^2 + [(\sin^2\phi\sin^2\theta + \cos^2\phi)/\gamma^2]}$$

Substituting $1/(1-(v^2/c^2))$ for $\gamma^2$ occurring in the expression under the square root, we get

$$C^+ = c\gamma^2 \sqrt{(\sin\phi\cos\theta - (v/c))^2 + (\sin^2\phi\sin^2\theta + \cos^2\phi)(1-(v^2/c^2))}$$

Simplifying the expression under the square root, we obtain

$$C^+ = c\gamma^2 (1 - (v/c)\sin\phi\cos\theta) \quad (6)$$

The above formula for the observed speed of light is in terms of the direction cosines observed by K. To convert the same in terms of direction cosines observed by K', we proceed as below.

Dividing both sides of equations (3), (4) and (5) by corresponding sides of equation (6), we get

$$\cos\theta' \sin\phi' = \frac{\sin\phi\cos\theta - \frac{v}{c}}{1 - \frac{v}{c}\sin\phi\cos\theta} \quad (7)$$

$$\sin\theta'\sin\phi' = \frac{\sin\theta\sin\phi}{\gamma\left(1-\frac{v}{c}\sin\phi\cos\theta\right)} \qquad (8)$$

and

$$\cos\phi' = \frac{\cos\phi}{\gamma\left(1-\frac{v}{c}\sin\phi\cos\theta\right)}. \qquad (9)$$

One may note that given $\theta$ and $\phi$, $\phi'$ can be evaluated from equation (9) and $\theta'$ can be evaluated from either equation (7) or (8).

From equation (7) we obtain (by treating ($\sin\phi\cos\theta$) as a single variable and solving for the same)

$$\sin\phi\cos\theta = \frac{\sin\phi'\cos\theta' + \frac{v}{c}}{1+\frac{v}{c}\sin\phi'\cos\theta'} \qquad (10)$$

Substituting for $\sin\phi\cos\theta$ from equation (10) into equation (6), we obtain

$$C^+ = \frac{c}{1+\frac{v}{c}\sin\phi'\cos\theta'} \qquad (11)$$

Equation (11) expresses the observed velocity of light in K' as a function of the observed direction cosines of the line of propagation in K'.

For any distance $\Delta L'$ in any arbitrary direction denoted by $\phi'$ and $\theta'$, the time taken will be

$$\Delta t' = \frac{\Delta L'}{C^+}$$

Substituting for $C^+$ from equation (11), we obtain

$$\Delta t' = \frac{\Delta L'}{c}\left(1+\frac{v}{c}\sin\phi'\cos\theta'\right)$$

$$= \frac{\Delta L'}{c} + \frac{v}{c^2}\Delta L'\sin\phi'\cos\theta'$$

Noting that $\Delta L' \sin\phi' \cos\theta' = \Delta x'$, we obtain

$$\Delta t' = \frac{\Delta L'}{c} + \frac{v}{c^2} \Delta x' \qquad (12)$$

In any closed path the summation of the second term vanishes and thus the average round-trip speed of light is observed to be c.

Further when we shift to the "proper" synchronization convention of K' given by

$$t'_E = t' - \frac{vx'}{c^2} \qquad (13)$$

we get, $\Delta t' = \Delta t'_E + \frac{v}{c^2} \Delta x' \qquad (14)$

Comparing equations (12) and (14), we get

$$\Delta t'_E = \frac{\Delta L'}{c} \qquad (15)$$

Therefore under the proper synchronization convention of K', not only the round-trip speed but also the one-way speed of light remains constant.

## 4. Discussion

The switch of the synchronization convention from that of K to K' as given by equation (13) can be traced to the discussion on the Relativity of Simultaneity [7]. Herein Einstein develops the definition for the statement that "two spatially separated events are simultaneous". In a hypothetical conversation he asks the reader to provide such a definition or method to determine whether two spatially separated events are simultaneous or not. After thinking the matter over for some time, the reader replies that if (in an inertial frame) two events occur at spatial locations A and B and an observer at the mid point M of the line joining AB, visually observes the two events at the same time, then these two events are simultaneous. Continuing his hypothetical conversation Einstein says that "your definition would certainly be right, if only I knew that the light traveled at the same speed along A → M and along B → M. But an examination of this supposition would only be possible if we already had at our disposal the means of measuring time. It would thus appear as though we were moving here in a logical circle." The reply from the hypothetical reader in Einstein's own words is as follows: "After further consideration you cast a somewhat disdainful glance at me – and rightly so – and you declare: ' I maintain my previous definition nevertheless, because in reality it assumes absolutely nothing about light. There is only *one* demand to be made of the definition of simultaneity, namely, that in every real case it must supply us with an empirical decision ……., that light requires the same time to traverse the path A → M as for the path B → M is in reality neither a *supposition* nor a *hypothesis* about the physical nature of light but a *stipulation*[1] which I can make of my own free will in order to arrive at a definition of simultaneity."

The above hypothetical conversation essentially means that the one-way speed of light in any path in a given inertial frame has to be assumed to be the average round-trip speed of light. If the light assumes different speeds in different directions, then in order to measure these we need to have a set of spatially synchronized clocks a-priori. A synchronization convention is required to develop such a set of clocks [5, 6]. The assumption of the one-way speed of light to be a constant within that inertial frame is the best possible convention for any given inertial frame [5, 7]. Thus observers in every inertial frame assume that light travels at same speed in all directions between co-moving objects of that inertial frame. Because of this, the synchronization of spatially separated clocks is unique to every inertial frame and the concept of relativity of simultaneity follows from these different synchronization conventions adopted by different inertial frames.

## 5. Conclusions

Under the IST transformation, the average speed of light in any closed path is constant. In any given line segment, the speed is given by equation (11). The Lorentz transformation is obtained from the IST transformation by switching to the 'proper' synchronization convention of the target inertial frame, K'. Under the Lorentz transformation, the speed of light remains constant in every segment of the path.

## Acknowledgements

The author acknowledges the useful interactions and inputs provided by Dr. G.M. Prabhu of Iowa State University in the development of this article.

The author further acknowledges the invaluable and thoughtful comments of the anonymous referee which have contributed to the improvement of this article.

---

[1] Other translations translate the word "*festsetzung*" as "convention" instead of "stipulation" [8].